\documentclass[twocolumn,showpacs,preprintnumbers,a4,prb]{revtex4}
\usepackage{graphicx}
\usepackage{dcolumn}
\usepackage{amsmath}
\usepackage{amssymb}
\usepackage{bm}

\begin{document}

\title{Enhancement of Superconductivity by Exchange Bias}
\author{D. Stamopoulos,\footnote[1]{Author to whom correspondence
should be addressed (densta@ims.demokritos.gr)} E. Manios, and M.
Pissas}

\affiliation{Institute of Materials Science, NCSR "Demokritos",
153-10, Aghia Paraskevi, Athens, Greece.}
\date{\today}

\begin{abstract}
In this work we study the transport properties of hybrids that
consist of exchange biased ferromagnets (FMs) combined with a
low-T$_c$ superconductor (SC). Not only different FMs but also
various structural topologies have been investigated: results for
multilayers of La$_{1-x}$Ca$_{x}$MnO$_{3}$ combined with Nb in the
form of
[La$_{0.33}$Ca$_{0.67}$MnO$_{3}$/La$_{0.60}$Ca$_{0.40}$MnO$_{3}]_{15}$/Nb,
and for more simple Ni$_{80}$Fe$_{20}$/Nb/Ni$_{80}$Fe$_{20}$
trilayers and Ni$_{80}$Fe$_{20}$/Nb bilayers are presented. The
results obtained in all hybrid structures studied in this work
clearly uncover that {\it the exchange bias mechanism promotes
superconductivity}. Our findings assist the understanding of the
contradictory results that have been reported in the recent
literature regarding the transport properties of relative FM/SC/FM
spin valves.
\end{abstract}


\maketitle

\section{Introduction}

Lately, it has been shown that the combination of ferromagnets
(FMs) with superconductors (SCs)
\cite{BuzdinReview,BergeretReview} offers an efficient way to
bypass the natural restrictions that exist in plain materials
regarding their electronic and magnetic properties. These FM/SC
hybrids are promising for the design of oncoming electronics since
the efficient magnetic modulation of the superconducting
properties could allow the direct implementation of novel
apparatus as FM/SC/FM spin valves.
\cite{Buzdin99,Tagirov99,Gu02,Potenza05,Moraru06,Pena05,Rusanov06}
Although these FM/SC/FM hybrids have attracted much interest
\cite{Buzdin99,Tagirov99,Gu02,Potenza05,Moraru06,Pena05,Rusanov06}
certain discrepancies that have been reported in the recent
literature relent the complete understanding of the underlying
physical mechanisms.

The present work offers results on the transport properties of
exchange biased FMs \cite{Schuller99} combined with a low-T$_c$
SC. Two quite different FMs, namely La$_{1-x}$Ca$_{x}$MnO$_{3}$
and Ni$_{80}$Fe$_{20}$, and also various structural topologies
have been investigated. More specifically, MLs of
La$_{1-x}$Ca$_{x}$MnO$_{3}$ combined with Nb in the form of
[La$_{0.33}$Ca$_{0.67}$MnO$_{3}$/La$_{0.60}$Ca$_{0.40}$MnO$_{3}]_{15}$/Nb,
and more simple Ni$_{80}$Fe$_{20}$/Nb/Ni$_{80}$Fe$_{20}$ trilayers
(TLs) and Ni$_{80}$Fe$_{20}$/Nb bilayers (BLs) are studied. The
aim of our study was to uncover the reasons responsible for the
contradictory results that have been reported very recently
regarding the transport properties of relative FM/SC/FM spin
valves. \cite{Gu02,Potenza05,Moraru06,Pena05,Rusanov06} {\it A
novel feature that is observed in the transport data obtained in
all different hybrids is that the exchange bias mechanism enhances
the SC's resistive transition}. This surprising experimental fact
together with the magnetization data that were reported very
recently \cite{StamopoulosPRB05,StamopoulosPRB06}, imply that {\it
the coexistence of exchange bias and superconductivity could have
a positive synergetic effect}.

\section{Preparation of samples and experimental details}

The ML/SC hybrids have the form
[La$_{0.33}$Ca$_{0.67}$MnO$_{3}$/La$_{0.60}$Ca$_{0.40}$MnO$_{3}]_{15}$/Nb
with [d$_{AF}=4$/d$_{FM}=4$]$_{15}$/d$_{SC}=100$ (in nm units). In
these samples a FM buffer layer with d$_{FM}=50$ nm has been used
since we expected that this should act as a main reservoir for
generating stray fields that, in addition to the exchange bias
mechanism, could influence the SC (see inset (c) in
Fig.\ref{b1}(a)). Information on the ML preparation may be found
in Ref.\onlinecite{StamopoulosPRB05}. The MLs exhibit Curie
critical temperature T$_c^{ML}=230$ K.

The FM/SC/FM TLs are constructed of
Ni$_{80}$Fe$_{20}$/Nb/Ni$_{80}$Fe$_{20}$ (see inset (c) in
Fig.\ref{b3}(a)) with d$_{FM}=19$/d$_{SC}=50$/d$_{FM}=19$ (in nm
units). Also, more simple Ni$_{80}$Fe$_{20}$/Nb BLs with
d$_{FM}=19$/d$_{SC}=50$ (in nm units) have been studied. For the
Ni$_{80}$Fe$_{20}$ (NiFe) layers rf-sputtering was employed at
$30$ W in $4$ mTorr Ar atmosphere ($99.999\%$ pure). We should
stress that: (i) all depositions were carried out {\it at room
temperature} and (ii) {\it no} magnetic field was applied during
the deposition of NiFe. In order to incorporate the exchange bias
only the NiFe {\it bottom} layers were treated by annealing (see
\cite{Schuller99} and references therein). Afterwards, the
complete TL and BL structures were deposited on top. The blocking
temperature \cite{Schuller99} of the NiFe bottom layers is
T$_B=100$ K. Information on the preparation of the SC may be found
in Ref.\onlinecite{Stamopoulos05PRB}. The SC critical temperature
of the produced films ranges in $7.8$ K$<$T$_c^{SC}<8.3$ K.

A commercial superconducting quantum interference device (SQUID)
(Quantum Design) was used as a host cryostat. In all transport
measurements the standard four-point configuration was used, the
external magnetic field was applied parallel to the hybrids and
the dc-current was normal to the external field (see insets (c) in
Fig.\ref{b1}(a) and Fig.\ref{b3}(a)).

\section{Experimental results}

\subsection{ML/SC hybrids}

\begin{figure}[tbp] \centering%
\includegraphics[angle=0,width=6.5cm]{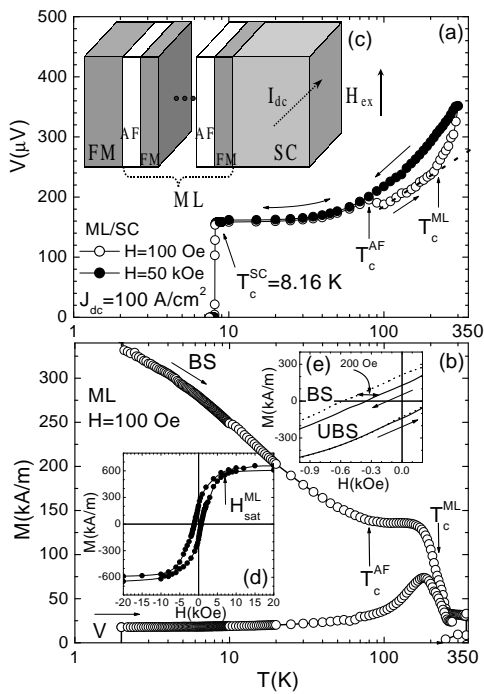}
\caption {(a) Voltage curve V(T) measured in a ML/SC hybrid from
below T$_c^{SC}=8.16$ K up to T$=350$ K for warming under an
external field $H_{ex}=100$ Oe when the ML was (V) virgin (open
circles) and when (BS) biased (solid circles). (b) Detailed
magnetization m(T) curves from T$=2$ K up to T$=350$ K for warming
under $H_{ex}=100$ Oe when the ML was V and BS. Inset (c) shows a
schematic representation of the ML/SC hybrid structure and of the
experimental configuration. Insets (d) and (e) show M(H) loops
obtained at T$=10$ K when the ML was BS and unbiased but initially
saturated (UBS). The ML saturates at H$_{sat}^{ML}=7$ kOe (inset
(d)) and the loop obtained under BS conditions is shifted by $200$
Oe when compared to the UBS one (inset (e)).}
\label{b1}%
\end{figure}%

Figures \ref{b1}(a) and \ref{b1}(b) show in semilogarithmic plots
representative transport and magnetization data that are needed
for a thorough characterization of the produced ML/SC hybrids. In
the lower panel we present detailed magnetization m(T) curves from
T$=2$ K up to T$=350$ K for the pure ML (prior to the deposition
of the Nb layer) under $H_{ex}=100$ Oe. The data were obtained
while the ML was warmed and for two distinct cases: when the ML is
initially virgin (V) and when biased (BS) [by cooling it from
T$>$T$_c^{ML}=230$ K under an external field $H_{ex}=50$ kOe that
it was lowered to $H_{ex}=100$ Oe when the desired temperature
T$=2$ K was achieved]. First of all, we notice that the very small
magnetization observed in the V curve in the low-temperature
regime clearly indicates that the FM layers are coupled
antiferromagnetically. The magnetization is almost constant up to
the N$\acute{e}$el critical temperature of the AF layers
T$_c^{AF}=80$ K. For T$>$T$_c^{AF}=80$ K the V curve increases
exhibiting a maximum at around T$=180$ K and subsequently gets
zero as T$_c^{ML}=230$ K is exceeded. The BS curve is also closely
related to the V one presenting distinct features at the
respective N$\acute{e}$el and Curie critical temperatures
T$_c^{AF}=80$ K and T$_c^{ML}=230$ K. As expected, above
T$_c^{ML}=230$ K the magnetization gets zero. For T$_c^{AF}=80$
K$<$T$<$T$_c^{ML}=230$ K the BS curve attains a constant value,
while for T$<$T$_c^{AF}=80$ K it exhibits high values. The
respective insets (d) and (e) show M(H) data for the ML obtained
at T$=10$ K. Inset (d) show a representative loop in an extended
field regime, while inset (e) focuses in the low-field regime for
the M(H) loops obtained when the ML was biased (BS) and unbiased
but initially saturated (UBS). These data clearly show that the
exchange bias mechanism \cite{Schuller99} controls the magnetic
behavior of the ML. The characteristic features observed in the
m(T) curves of the lower panel are also visible in the respective
V(T) ones presented in the upper panel. Its inset (c) shows a
schematic representation of the ML/SC hybrid structure and of the
measuring configuration. As may be seen the resistive transition
of the SC occurs at T$_c^{SC}=8.16$ K.

\begin{figure}[tbp] \centering%
\includegraphics[angle=0,width=6cm]{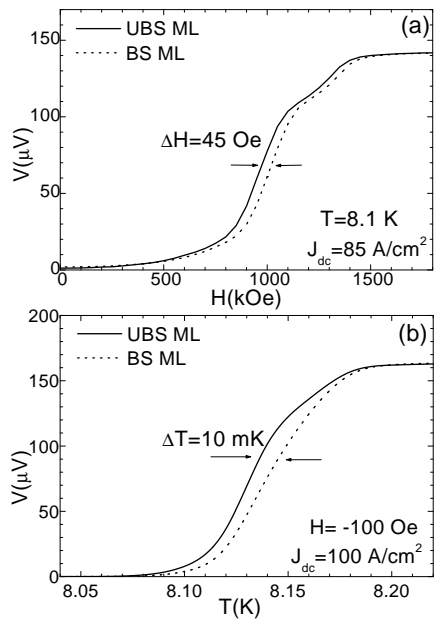}
\caption {Representative voltage curves obtained while lowering
either (a) the applied magnetic field at T$=8.1$ K or (b) the
temperature at $H_{ex}=-100$ Oe. Dotted curves were obtained when
the ML was BS, while solid ones when it was UBS.}
\label{b2}%
\end{figure}%

The detailed magnetization data that were presented in Refs.
\onlinecite{StamopoulosPRB05,StamopoulosPRB06} for the same ML/SC
hybrids revealed that the exchange bias mechanism influences the
SC's magnetic behaviour. In order to investigate if, except for
the magnetic behavior, the exchange bias could influence the SC's
transport properties we performed resistivity measurements.
Representative data are shown in Figs.\ref{b2}(a) and \ref{b2}(b)
revealing that the exchange bias mechanism clearly affects the
current-carrying capability of our ML/SC hybrids. Measurements
were performed while lowering either the applied magnetic field
(upper panel) or the temperature (lower panel) for two different
conditions: when the ML was initially BS (dotted curves) and when
UBS (solid curves) [see also inset (e) in Fig.\ref{b1}(b)]. We
observe that the BS curves are placed below the UBS ones. This is
a clear evidence {\it that superconductivity is enhanced when the
SC experiences the exchange bias imposed by the adjacent ML}.

\subsection{FM/SC/FM TL and FM/SC BL hybrids}

\begin{figure}[tbp] \centering%
\includegraphics[angle=0,width=6.5cm]{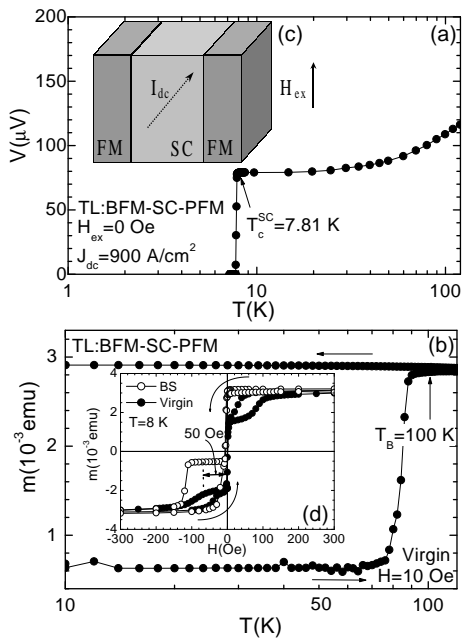}
\caption {(a) Voltage curve V(T) measured in a TL from
T$_c^{SC}=7.81$ K up to T$=120$ K for $H_{ex}=0$ Oe. (b) Zero
field cooled and field cooled magnetization m(T) curves from
T$=10$ K up to T$=120$ K under $H_{ex}=10$ Oe when the TL is
virgin. Inset (c) shows a schematic representation of the FM/SC/FM
TL hybrid and of the experimental configuration. Inset (d)
presents isothermal m(H) curves obtained at T$=8$ K when the TL
was virgin (solid circles) and BS (open circles). The loop
obtained under BS condition is shifted by $50$ Oe when compared to
the virgin one.}
\label{b3}%
\end{figure}%

In order to investigate the generic character of this result we
examined another quite different FM constituent, namely NiFe
combined in two NiFe/Nb/NiFe TL and NiFe/Nb BL topologies with the
SC. Figures \ref{b3}(a) and \ref{b3}(b) show in semilogarithmic
plots representative results that reveal the existence of exchange
bias in a NiFe/Nb/NiFe TL. We remind that only the bottom NiFe
layer exhibits exchange bias so that it is called BFM ("B" stands
for biased), while the as-deposited top NiFe layer is noted as PFM
("P" stands for plain). Panel (a) presents the resistive curve
from below T$_c^{SC}=7.81$ K up to T$=120$ K, while in panel (b)
presented are the zero field cooled and field cooled m(T) curves
under $H_{ex}=10$ Oe when the sample is virgin. Inset (d) of panel
(b) presents m(H) loops obtained at T$=8$ K$>$T$_c^{SC}=7.81$ K
when the TL was virgin and BS [biasing is achieved by cooling the
TL from T$>$T$_B=100$ K down to T$=8$ K under an external field
$H_{ex}=500$ Oe].

\begin{figure}[tbp] \centering%
\includegraphics[angle=0,width=6.5cm]{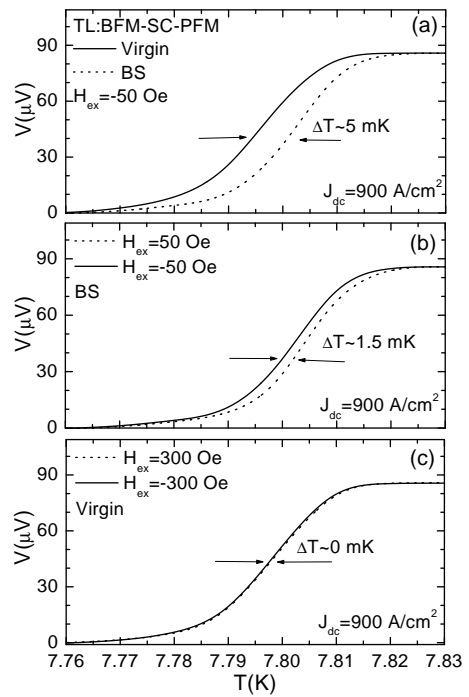}
\caption {Representative voltage curves as a function of
temperature (a) at $H_{ex}=-50$ Oe when the TL is virgin and BS,
(b) at $H_{ex}=-50$ and $50$ Oe when it is BS, and (c) at
$H_{ex}=-300$ and $300$ Oe when it is virgin.}
\label{b4}%
\end{figure}%

Measurements as a function of temperature for different magnetic
states of the TL are shown in Figs.\ref{b4}(a)-\ref{b4}(c). In
nice agreement with the ones presented in Fig.\ref{b2} for the
ML/SC hybrid, these data reveal that {\it the exchange bias
clearly promotes the resistive critical temperature of the TL} as
for instance may be seen in panel (a). In panel (b) although both
curves were obtained for BS TL they refer to different applied
fields, $H_{ex}=-50$ and $50$ Oe. Thus, these measurements refer
to different magnetization configuration of the outer NiFe layers:
at $H_{ex}=-50$ Oe the layers are antiparallel and at $H_{ex}=50$
Oe they are parallel (see the BS loop in inset (d) of
Fig.\ref{b3}(b)). This information will help us to discuss the
recent discrepancies that have been reported
\cite{Gu02,Potenza05,Moraru06,Pena05,Rusanov06} in relevant
FM/SC/FM spin valves. Since in all measurements presented in
Figs.\ref{b2}(b) and \ref{b4}(a)-\ref{b4}(b) the observed
temperature shift is very small (but, at least, comparable to the
ones presented in
Refs.\onlinecite{Gu02,Potenza05,Moraru06,Pena05,Rusanov06}) we
performed some test measurements in order to ensure that these
shifts are motivated by the physics of the studied systems and are
not coincidental. The results of panel (c) serve this aim since
they clearly show that at the symmetric points $H_{ex}=-300$ and
$300$ Oe of the virgin TL the obtained V(T) curves clearly
coincide as they should (see the virgin loop in inset (d) of
Fig.\ref{b3}(b)).

\begin{figure}[tbp] \centering%
\includegraphics[angle=0,width=6.5cm]{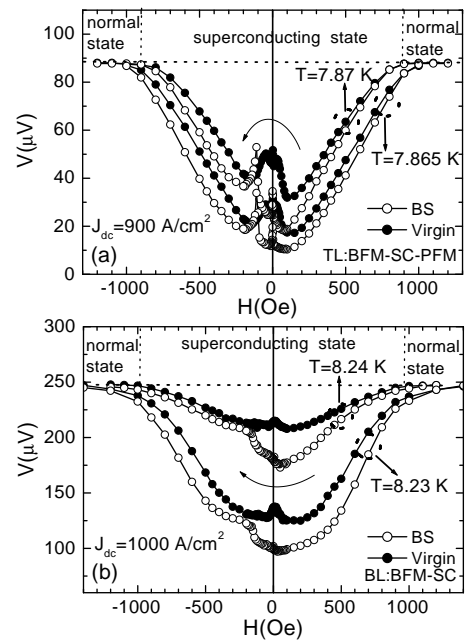}
\caption {Representative isothermal voltage curves, V(H) for (a) a
TL and (b) a BL obtained close to their T$_c^{SC}=7.89$ K and
T$_c^{SC}=8.26$ K, respectively. Both BS (open circles) and virgin
(solid circles) curves are presented.}
\label{b5}%
\end{figure}

Figure \ref{b5}(a) shows V(H) curves for a second TL obtained
close to its T$_c^{SC}=7.89$ K. Both virgin and BS curves are
shown. Firstly, the virgin curves present a broad peak around zero
field. As we see from the virgin m(H) curve in inset (d) of
Fig.\ref{b3}(b) at this field region the TL's magnetization
gradually reverses. This result is identical to the one presented
by V. Pe\~{n}a et al. in Ref.\onlinecite{Pena05}. {\it More
importantly, here we clearly demonstrate that this peak may be
almost entirely suppressed by the application of the exchange
bias.} Figure \ref{b5}(b) shows the respective V(H) curves for a
BL obtained close to its T$_c^{SC}=8.26$ K. In both the TL (panel
(a)) and the BL (panel (b)) we see that while in the normal state
the virgin and BS curves coincide, as we enter in the
superconducting state these curves diverge, with {\it the BS curve
placed significantly below the virgin one} \cite{MR}.

\section{Discussion}

Let us now discuss the contradictory experimental results
\cite{Gu02,Potenza05,Moraru06,Pena05,Rusanov06} that have been
reported in the recent literature for relevant spin valves. The
concept of a SC spin valve is based on a FM/SC/FM TL structure
\cite{Buzdin99,Tagirov99} where the nucleation of
superconductivity can be controlled at will by the {\it relative}
magnetization configuration of the outer FM layers. J.Y. Gu et al.
\cite{Gu02} were the first who reported on the experimental
realization of a
[Ni$_{82}$Fe$_{18}$-Cu$_{0.47}$Ni$_{0.53}$]/Nb/[Cu$_{0.47}$Ni$_{0.53}$-Ni$_{82}$Fe$_{18}$]
spin valve. In that work \cite{Gu02} the exchange bias mechanism
was used since an additional Fe$_{50}$Mn$_{50}$ layer was
introduced in order to control the magnetization of the one FM
layer. It was observed that when the magnetizations of the two FM
layers were {\it antiparallel (parallel)} the resistive transition
of the SC was placed at {\it higher (lower)} temperatures.
\cite{Gu02} Soon after, A. Potenza and C.H. Marrows
\cite{Potenza05}, and also I.C. Moraru et al. \cite{Moraru06} by
employing also the exchange bias mechanism confirmed the results
of Ref. \onlinecite{Gu02}. Contrary to those results
\cite{Gu02,Potenza05,Moraru06}, in the experiments of V. Pe\~{n}a
et al. \cite{Pena05} and of A.Yu. Rusanov et al. \cite{Rusanov06},
where
La$_{0.7}$Ca$_{0.3}$MnO$_3$/YBa$_2$Cu$_3$O$_7$/La$_{0.7}$Ca$_{0.3}$MnO$_3$
and Ni$_{80}$Fe$_{20}$/Nb/Ni$_{80}$Fe$_{20}$ TLs were studied
respectively, the opposite behavior was observed. Both works
\cite{Pena05,Rusanov06} reported that the {\it antiparallel
(parallel)} magnetization configuration of the FM layers {\it
suppresses (enhances)} superconductivity. We stress that in Refs.
\onlinecite{Pena05,Rusanov06} the exchange bias mechanism was {\it
not} employed. According to our results when the SC experiences
the exchange bias its resistive transition is enhanced. This
finding could explain the discrepancies reported in
Refs.\onlinecite{Gu02,Potenza05,Moraru06,Pena05,Rusanov06}. The
results presented in Fig.\ref{b4}(b) reveal that the {\it
antiparallel} magnetization configuration of the outer NiFe layers
{\it suppresses} superconductivity in agreement with
Refs.\onlinecite{Pena05,Rusanov06}.

Finally, exchange bias \cite{Schuller99} is related to
unidirectional magnetic anisotropy and phenomenologically it can
be viewed as a parameter that controls the nucleation and
orientation of magnetic domains in a FM; a key parameter for the
FM/SC/FM spin valves. Since we observed that the parallel magnetic
configuration enhances superconductivity we may assume that
exchange bias also promotes superconductivity by promoting the
alignment of the magnetic domains over the whole FM. Ultimately,
this could somehow be related to the existence of a spin-triplet
supercurrent \cite{Keizer06,Sosnin06} since, according to basic
knowledge under such conditions a spin-singlet one should be
seriously suppressed due to the detrimental influence of the FM's
exchange field. \cite{BergeretReview} More specific experiments
are needed to confirm this hypothesis.

\section{Conclusions}

Summarizing, we presented transport data in
[La$_{0.33}$Ca$_{0.67}$MnO$_{3}$/La$_{0.60}$Ca$_{0.40}$MnO$_{3}]_{15}$/Nb,
Ni$_{80}$Fe$_{20}$/Nb/Ni$_{80}$Fe$_{20}$ and Ni$_{80}$Fe$_{20}$/Nb
hybrids. A novel finding that we observed is that in all different
structures the exchange bias mechanism improves the resistive
transition of the FM/SC hybrid. Since this feature is independent
of the specific FM material and of the structural topology we
believe that it should be generic in all FM/SC hybrids. Our
experiments assist the understanding of the discrepancies that
have been reported in the recent literature regarding relative
FM/SC/FM spin valves.

\begin{acknowledgments}
Dr. N. Moutis is acknowledged for valuable help during the
preparation of samples.
\end{acknowledgments}

\pagebreak


\begin{references}

\bibitem{BuzdinReview} A.I. Buzdin, Rev. Mod. Phys. {\bf 77}, 935
(2005).

\bibitem{BergeretReview} F.S. Bergeret, A.F. Volkov, and K.B. Efetov, Rev. Mod. Phys. {\bf 77}, 1321
(2005).

\bibitem{Buzdin99} A.I. Buzdin, A.V. Vedyayev, and N.V. Ryzhanova, Europhys. Lett. {\bf 48}, 686
(1999).

\bibitem{Tagirov99} L.R. Tagirov, Phys. Rev. Lett. {\bf 83}, 2058 (1999).

\bibitem{Gu02} J.Y. Gu, C.-Y.You, J.S. Jiang, J. Pearson, Ya.B. Bazaliy, and S.D. Bader,
Phys. Rev. Lett. {\bf 89}, 267001 (2002).

\bibitem{Potenza05} A. Potenza, and C.H. Marrows, Phys. Rev. B {\bf 71}, 180503(R) (2005).

\bibitem{Moraru06} I.C. Moraru, W.P. Pratt, Jr., and N.O. Birge, Phys. Rev. Lett. {\bf 96}, 037004
(2006).

\bibitem{Pena05} V. Pena, Z. Sefrioui, D. Arias, C. Leon, J. Santamaria, J.L. Martinez,
S.G.E. te Velthuis, and A. Hoffmann, Phys. Rev. Lett. {\bf 94},
57002 (2005).

\bibitem{Rusanov06} A.Yu. Rusanov, S. Habraken, and J. Aarts, Phys. Rev. B {\bf 73}, 060505(R)
(2006).

\bibitem{Schuller99} J. Nogues, and I. K. Schuller, J. Magn. Magn. Mater. {\bf 192}, 203 (1999).

\bibitem{StamopoulosPRB05} D. Stamopoulos, N. Moutis, M. Pissas, and D. Niarchos, Phys. Rev. B. {\bf 72},
212514 (2005).

\bibitem{StamopoulosPRB06} D. Stamopoulos, and M. Pissas, Phys. Rev. B. {\bf 73}, 132502 (2006).

\bibitem{Stamopoulos05PRB} D. Stamopoulos, M. Pissas, and E. Manios, Phys. Rev. B {\bf 71}, 014522 (2005).

\bibitem{MR} The resistive peaks in the BS curves
relate to the reversal of each NiFe layer's magnetization. This
effect will be discussed elsewhere.

\bibitem{Keizer06} R.S. Keizer, S.T.B. Goennenwein, T.M. Klapwijk, G. Miao, G. Xiao, and A. Gupta, Nature {\bf 439}, 825 (2006).

\bibitem{Sosnin06} I. Sosnin, H. Cho, V.T. Petrashov, and A.F. Volkov, Phys. Rev. Lett. {\bf
96}, 157002 (2006).

\end{references}
\end{document}